\documentclass[12pt,preprint]{aastex}

\usepackage{amsmath}

\begin{document}

\title{A Strange Star Scenario for the Formation of Eccentric Millisecond Pulsar/Helium White Dwarf Binaries}

\author{Long Jiang$^{1}$, Xiang-Dong Li$^{1,2}$, Jishnu Dey$^3$, and Mira Dey$^3$}

\affil{$^{1}$Department of Astronomy, Nanjing University, Nanjing
210046, China}

\affil{$^{2}$Key laboratory of Modern Astronomy and Astrophysics
(Nanjing University), Ministry of Education, Nanjing 210046, China;
lixd@nju.edu.cn}

\affil{$^{3}$Department of Physics, Presidency University, 86/1,
College Street, Kolkata 700 073, India; jishnu.dey@gmail.com,
mira.dey@gmail.com}

\begin{abstract}
According to the recycling scenario, millisecond pulsars (MSPs) have
evolved from low-mass X-ray binaries (LMXBs). Their orbits are
expected to be circular due to tidal interactions during the binary
evolution, as observed in most of the binary MSPs. There are some
peculiar systems that do not fit this picture. Three recent examples
are PSRs J2234$+$06, J1946$+$3417 and J1950$+$2414, all of which are
MSPs in eccentric orbits but with mass functions compatible with
expected He white dwarf companions. It has been suggested these MSPs
may have formed from delayed accretion-induced collapse of massive
white dwarfs, or the eccentricity may be induced by dynamical
interaction between the binary and a circumbinary disk. Assuming
that the core density of accreting neutron stars in LMXBs may reach
the density of quark deconfinement, which can lead to phase
transition from neutron stars to strange quark stars, we show that
the resultant MSPs are likely to have an eccentric orbit, due to the
sudden loss of the gravitational mass of the neutron star during the
transition. The eccentricities can be reproduced with a reasonable
estimate of the mass loss. This scenario might also account for the
formation of the youngest known X-ray binary Cir X$-$1, which also
possesses a low-field compact star in an eccentric orbit.
\end{abstract}


\section{Introduction}
Millisecond pulsars (MSPs) are low-field ($B\sim 10^{8}-10^{9}$ G),
rapidly rotating neutron stars (NSs) with spin periods
${P}_\mathrm{s}\lesssim 30$ ms \citep{Lorimer2008}. In the standard
recycling scenario \citep{Alpar1982,Radhakrishnan1982,BH1991},
binary MSPs are regarded as the descendants of low-mass X-ray
binaries (LMXBs), in which a NS has accreted sufficient mass from
its companion star, and spun up to millisecond periods. Recent
observations of the LMXB/MSP transition in PSRs J1023$+$0038
\citep{Archibald2009}, J1824$-$2452I \citep{Papitto2013}, and
J1227$-$4853 \citep{Roy2015} have lent strong support to the
evolutionary link between LMXBs and MSPs.

If the initial orbital period of a LMXB is longer than the so-called
bifurcation period \citep[e.g.,][]{Pylyser1989}, the mass transfer
begins when the companion star evolves to be a (sub)giant. The
binary orbit continues expanding until the donor star loses most of
its hydrogen envelope, forming a low-mass He white dwarf (WD). Since
the mass transfer timescale is much longer than tidal
circularization timescale, the orbits of binary MSPs are highly
circularized \citep{Phinney1992}, except those in globular clusters
where the orbits may be perturbed by dynamical interactions in the
dense stellar environment.

The discovery of PSR J1903$+$0327 presents a challenge to the
recycling scenario, which is a binary MSP with a high eccentricity
($e\sim 0.44$) \citep{Champion2008}. Located in the Galactic field,
this pulsar is accompanied by a G-type main-sequence secondary. It
was suggested to stem from a hierarchical triple system
\citep{Freire2011,Portegies2011,Pijloo2012}, or be a newborn NS
experienced supernova fallback disk accretion \citep{Liu2009}.
%Two more recent examples are PSRs J2234$+$06 \citep{Deneva2013} and
%J1946$+$3417 \citep{Barr2013}. Both are Galactic field MSPs with a
%He WD (of mass around $0.24M_{\odot}$ for a $1.4M_{\sun}$ NS) in an
%around 30 day orbit. The short spin periods ($P\simeq 3$ ms)
%indicate that they have experienced extensive mass transfer
%processes during the LMXB evolution, but the eccentric orbits (with
%$e\simeq 0.13$) are difficult to understand.
Since then, some more eccentric MSPs have been discovered, including
PSRs J2234$+$06 \citep{Deneva2013}, J1946$+$3417 \citep{Barr2013}
and J1950$+$2414 \citep{Knispel2015}\footnote{Another MSP with an
eccentric orbit will be presented in Camilo et al. (2015, in
preparation). A previously known system, PSR J1618$-$3919 might also
be related: it has a spin period of 12 ms and an orbital
eccentricity of 0.027 \citep{Bailes2010}.}. All of these objects are
Galactic field MSPs with a He WD (of mass around $0.24 M_{\odot}$
for a $1.4 M_{\odot}$ NS) in orbits with periods ranging from 22 to
32 days. The short spin periods ( $2 \lesssim P_{\rm s} \lesssim 12$
ms) indicate that they have experienced extensive mass transfer
processes during the LMXB phase, but the eccentric orbits (with
$0.027 < e < 0.13$) are difficult to understand in standard LMXB
evolution scenarios. Like in the case of PSR J1903$+$0327, it is
still in principle possible that all these systems formed by the
disruption of a triple system. However, the strong similarity in the
orbital properties of all the recently discovered pulsars is not
what one would expect from the disruption of a triple system: this
is a chaotic process that should produce a wide range of orbital
periods and orbital eccentricities. The close similarity of these
systems suggests instead a common evolutionary mechanism, different
from what formed PSR J1903$+$0327.

\cite{Freire2014} suggested that these MSPs formed from delayed
accretion-induced collapse of massive WDs. According to their
scenario, the progenitor binary may consist of a $7M_{\odot}$
primary and a $2M_{\odot}$ secondary. The primary star evolves to be
an ONeMg WD of mass $1.2M_{\sun}$ while the secondary is still on
its main sequence stage. In the following evolution, the secondary
fills its Roche lobe and transfers hydrogen-rich material to the WD,
so that the mass of WD increases to the Chandrasekhar limit.
However, if the WD is fast spinning, it may keep accreting without
collapsing to be a NS \citep[e.g.][]{Yoon2004}. The WD-NS conversion
only occurs when the WD has spun down and the centrifugal forces
that sustain the star are weakened, which might take several Gyr.
Then the sudden released gravitational binding energy during the
collapse increases the orbital period and imposes an eccentricity to
the former circularized orbit. The re-circularization times for such
MSP/WD binaries are generally much longer than Hubble time
\citep{Zahn1977,Hut1981}.
%A caveat for this scenario is that at the
%late stage of mass transfer, the accretion rate of the WD is usually
%too low to keep stable hydrogen burning on the surface of the WD, so
%that almost all of the accreted matter may be lost from the WD
%\citep[e.g.,][]{Yaron2005}.
According to the calculation by \citet{Tauris2013}, there is
an optimal range of orbital periods where the mass transfer  is both
dynamically stable (i.e., without entering common envelope
evolution) and rapid enough for stable thermonuclear burning of
Hydrogen \citep[e.g.,][]{Yaron2005}. This range of orbital periods
goes from 10 to 60 days, and the middle of this range is right where
the aforementioned systems are.
%\footnote{According to \cite{Yoon2004}, a massive WD can
%keep differential rotation only with the accretion rate higher than
%$3\times10^{-7}M_{\odot}\text{yr}^{-1}$.}.

Alternatively, \cite{Antoniadis2014} proposed that the
eccentricities of the two MSPs may be caused by dynamical
interaction between the binary and a circumbinary disk, which formed
from the material escaping the donor star during hydrogen-shell
flashes shortly before the WD cooling phase. Adopting a linear
perturbation theory following \cite{Dermine2013}, the author argued
that, a disk with a fine-tuned lifetime as long as $\sim10^5$ yr and
mass around a few $10^{-4}M_\odot$ may result in eccentricities of
$e\simeq 0.01 - 0.15$ for orbital periods between 15 and 50 days.

Adopting the idea that the orbital eccentricity is caused by sudden
mass loss, here we suggest another possible scenario to this
problem. During the recycling phase the NS accretes mass and angular
momentum from the donor star. When its central density goes up to
the critical density of quark deconfinement, a phase transition from
a NS to a strange quark star (SS) may take place within a short
time, and the loss of binding energy of the NS, if not big enough to
disrupt the binary, can induce an eccentricity of the orbit. If the
phase transition occurs at the end of or after the recycling
process, the SS appears as an eccentric binary MSP with a He WD
companion.

%In the following section we described the evolution code of binary;
%Simulated the AIC model and analyzed the result in section 3 we
%shown its difficulty to form such binary;  We discussed the state
%transition model in detail in section 4 and the parameter space of
%initial orbital period and donor mass is given by the simulation
%binary evolution; finally, in section 5 we discussed the possibility
%that these binaries stemmed from hierarchical triple systems which
%ejected one of their members and the detectable difference between
%these models are also given there.

\section{The NS-SS transition scenario}

Below we describe in some detail how the the NS-SS transition can
influence the formation of various types of MSPs.

The concept of SSs has been suggested decades ago
\citep[]{Itoh1970,Bodmer1971,Witten1984,Farhi1984,Haensel1986,
Alcock1986}, and it was suggested that, due to the absolute
stability of quark matter, at least some of the pulsars could be
SSs. Phase transition occurs when the central density of a NS rises
above the critical density for dissolution of baryons into their
quark constituents. The detailed processes of quark deconfinement in
massive NSs are not precisely known. It might be gradual, lasting
around $10^8$ yr \citep[]{Olinto1987,Horvath1988}, or in a
detonation mode, accompanied with energy released compatible with a
core collapse supernova \citep[e.g.,][]{Cheng1996,Quyed2002}. Based
on the standard equation of state (EOS) of neutron-rich matter,
\cite{Staff2006} pointed out that the critical density of quark
deconfinement is $\sim5\rho_0$, where
$\rho_0\sim2.7\times10^{14}\text{g cm}^{-3}$ is the nuclear
saturation density. Although different EOSs give different masses
when the critical density is reached, a gravitational mass of
$1.8M_{\odot}$ may be a reasonable estimate for a NS with a core of
quark matter \citep{Akmal1998}.

Quite a few specific SS candidates \citep[e.g., SAX
J1808.8$-$3658,][]{Li1999} have been proposed \citep[see][for a
review]{Weber2005}, usually based on their observational features in
X-rays. The formation of a SS can also impact the dynamical
evolution of the host binary, which is seldom mentioned in the
literature. We note that efficient mass growth of NSs is most likely
to take place in LMXBs\footnote{NSs in intermediate-mass X-ray
binaries usually accretes at a super-Eddington rate, so that most of
the mass may be ejected. During the subsequent LMXB evolution when
the accretion rate becomes sub-Eddington, the NSs can keep the
accreted material and grow in mass.}, and the phase transition may
occur in the following two situations.

In the first one, an accreting NS converts to be a SS during the
LMXB stage when it becomes sufficiently massive. Some baryonic mass
is abruptly transformed into its binding energy during the phase
transition, and the orbit expands and becomes eccentric. The donor
star is then detached from its Roche-lobe, temporarily halting the
mass transfer, and the SS becomes a MSP. If the orbit is close
enough, its high-energy radiation and particles might be able to
ablate the companion star, leading to the formation of redbacks
\citep[][for a review]{Roberts2013}. Otherwise, after some time the
companion star evolves to refill its Roche lobe and resumes the mass
transfer. The orbit is then invariably circularized by tidal
torques, so the binary evolves as a normal LMXB, producing a
circular MSP/He WD binary.

In the second, if the NS's spin is accelerated to be too fast, the
central density might be centrifugally diluted and lie below the
critical density for pure quark matter. Phase transition is delayed,
and takes place during the terminal stages of the mass transfer or
even the subsequent MSP phase, during which the NS is spun down due
to the propeller mechanism \citep{Tauris2012}, or magnetic dipole
and/or gravitational radiation, respectively. The binary will keep
its eccentricity because of the extremely long circularization time
for two compact component stars. That may be responsible for the the
eccentric orbits of PSRs J2234$+$06 and J1946$+$3417.

\section{Phase transition-induced orbital change}

Similar as the \cite{Freire2014} model, here the orbital
change in the MSP binaries is assumed to be due to instantaneous
mass loss. To find out the possible parameter space of the WD mass
and the mass loss during the phase transition for PSRs J2234$+$06
and J1946$+$3417 from the MSP mass, the orbital period, and the
eccentricity, we use a Monte Carlo method to simulate the orbital
change in the following steps.

{\noindent 1. We assume that the  current mass  $M_\text{SS}$ of the
MSP is randomly distributed  in the range of $1.4-2.0M_\odot$
\citep{g13}. The upper limit is set in accord with recent
measurements of two massive pulsars, PSR J1614$-$2230 of mass
$1.97(\pm0.04)M_\odot$ \citep{Demorest2010} and PSR J0348$+$0432 of
mass $2.01(\pm0.04)M_\odot$ \citep{Antoniadis2013}. }

{\noindent 2. The current mass $M_{\rm WD}$ of the He WD  is then
calculated from $M_\text{SS}$ and the observed mass function
$M^3_\text{WD}\text{sin}^3i/(M_\text{SS}+M_\text{WD})^2=0.0027$ for
PSR J2234$+$06 \citep{Deneva2013}, where $i$ is the inclination
angle of the orbit. We assume that $\cos i$ is uniformly distributed
between 0 and 1.

{\noindent 3. Assuming that the binary orbit is initially circular
and the current eccentricity (assumed to be in the range of
$0.11-0.15$) is induced by instantaneous gravitational mass loss of
the NS, we can get the amount of the mass loss $\Delta
M=e\cdot(M_\text{WD}+M_\text{SS})$ and the gravitational mass
$M_\text{NS}$ ($=M_\text{SS}+\Delta M$) of the NS \citep{BH1991}, if
the energy releasing during the phase transition is spherically
symmetric and there is no kick imparted to the SS.}

{\noindent 4. Not all of the obtained values of $\Delta M$ and
$M_{\rm WD}$ are physically reasonable, and we select those which
satisfy the following condition predicted by LMXB evolution. On one
hand, from $M_\text{SS}$,  $M_{\rm WD}$ and the current orbital
period (assumed to be in the range of $22-32$ days), we get the
orbital separation $a$. Combining $a$ and the eccentricity $e$
yields the separation $a_{\rm i}$ before the phase transition via
the equation $a_{\rm i}=a(1-e)$. On the other hand, we derive the
orbital period $P_\text{orb, i}$ before the phase transition from
$M_{\rm WD}$, using the orbital period - WD mass relation given by
\cite{TS1999} for solar metallicities ($Z=0.02$). Then from
$P_\text{orb, i}$, $M_\text{NS}$, and $M_\text{WD}$ we also obtain
$a_{\rm i}$ from the Kepler's law. We require that the value of this
$a_{\rm i}$ should match the former one within $1\%$.

The simulated $\Delta M$ vs. $M_{\rm WD}$ distribution is shown in
Fig.~1. The allowed maximum NS mass is around $2.3M_\odot$. It is
seen that the mass loss is about $0.20-0.34M_\odot$, and the WD mass
is constrained to be about $0.257-0.275M_\odot$ (red dots). If we
decrease $e$ down to 0.027 (the eccentricity of PSR J1618$-$3919,
which might have the same formation mechanism),  the mass loss can
be as low as about $0.05M_\odot$, while the upper limit of the
possible WD mass increases to about $0.278M_\odot$ (blue dots).
Simulation with the mass function $=0.0039$ for PSR J1946$+$3417
\citep{Barr2013} gives similar result, as shown in Fig.~2. These
numbers are compatible with the standard evolution of LMXBs.

The observed range of the orbital periods for eccentric MSPs
thus far is $22-32$ days, and the corresponding orbital periods
before phase transition are found to range from 15 to 35 days. It is
not clear why phase transition would only occur for such a
particular range of orbital periods. It might be related to the
requirement that phase transition (supposedly in a massive NS)
should occur at (or near) the end of mass transfer, in order to
produce observable eccentric MSPs. According to the studies of
\cite{Li1998} and \cite{TS1999} (see also Zhu et al. 2013), the
amount of mass accreted by a NS during the LMXB evolution generally
decreases with the orbital period, probably due to the
thermal-viscous instability in the accretion disks
\citep{Lasota2001}. In systems with longer orbital periods, the NS
may not accrete sufficient mass to trigger phase transition, while
NSs in narrower orbits may get enough material and undergo phase
transition during the LMXB phase, but the resulting eccentric orbits
would be circularized by the subsequent mass transfer. So we
tentatively speculate that only LMXBs with a particular range of
orbital periods may lead to the formation of eccentric MSPs.
However, we need to caution that the accretion processes in LMXBs
are rather complicated. Considering the fact the mass transfer is
very likely to be nonconservative and the big uncertainties in the
accretion efficiency, it is premature to accurately estimate the
accreted mass by a NS.

\section{Discussion}

Based on the assumption that massive NSs may convert to be SSs, we
suggest that the sudden loss of gravitational mass during the phase
transition may account for the eccentric orbits in the two recycled
binary pulsars, which are otherwise thought to be in circular
orbits.

%For the formation possibility from  triple system disruption we
%refer to the study of \cite{Portegies2011}. To resolve the formation
%problem of eccentric orbit MSP J1903+0327 they simulated the channel
%of triple system disruption. Majority of their simulation result to
%disruption of triple. While in that study the attention is payed to
%the ratio of the ejection of the secondary (WD, donor star of
%accreted) here the ejected object is the tertiary ( main-sequence,
%in the outer orbit ) and the ratio is $\sim0.3-\sim0.5$.

Recent population synthesis calculation by \citet{Zhu2013} suggested
that about $0.1-10\%$ of LMXBs can produce SSs, depending on the
masses of the nascent NSs and the fraction of transferred matter
accreted by the NSs. Observational clues have also been proposed for
the existence of SSs. For example, it is known that special global
oscillation modes (r-modes) that emit gravitational waves would
prohibit pulsars from spinning with high frequencies, unless the
damping of these modes, determined by the microscopic properties of
matter, can prevent this \citep[e.g.,][]{Andersson2001}.
\citet{Alford2014} showed that for each form of matter there is a
distinct region in the spin frequency/spindown-rate diagram where
r-modes can be present. They found that stars containing ungapped
quark matter are consistent with both the observed radio and X-ray
data, whereas, even when taking into account the considerable
uncertainties, NS models with standard viscous damping are
inconsistent with both data sets and additional damping mechanisms
would be required.

Our phase transition scenario can be compared with others in the
following aspects, which are to be tested by future observations.

\noindent 1. Although phase transitions are expected to occur in
massive NSs, the masses of the resultant SS MSPs could lie in a wide
range \citep{g13}, depending on the amount of the mass that is
transferred into the binding energy. This is similar to the
\cite{Antoniadis2014} prediction but different from the simplest
scenario in \citet[]{Freire2014} (that of a rigidly rotating WD
before the collapse), which predicts that these MSPs should be
exceptionally light.

\noindent 2. The eccentric MSPs probably have a He WD
companion with mass in a relatively narrow range ($\sim
0.25-0.28\,M_{\sun}$). With such WD companions, the NSs are likely
to be heavy at the moment of phase transition, implying that the
resultant SS pulsars are fully recycled pulsars.

\noindent 3. The eccentric MSPs may have kinematic
properties (i.e., Galactic heights and peculiar velocities) at least
similar as what is observed in the normal MSP population, since the
supernova that formed the NS should give the system a reasonable
kick. This is again similar to the \cite{Antoniadis2014} prediction
but contrasts with that of \citet[]{Freire2014}. In the latter
scenario, the AIC is the only supernova that ever occurs and,
because this supernova is so symmetric, the eccentric systems
generated would have very small peculiar velocities and Galactic
heights.

As noted before, phase transition may also occur during the LMXB
phase. The conversion of a NS into a SS could lead to a quark nova
\citep{Quyed2002,Quyed2011}, with energy released comparable to a
normal supernova. A peculiar X-ray binary, Cir X$-$1, might be
produced in this way. This source has an orbital period of 16.6 days
\citep{Kaluzienski1976}. It does not seem to fit either a high-mass
or a low-mass X-ray binary. Detection of type-I X-ray bursts
\citep{Tennant1986,Linares2010} firmly establishes its nature as a
low-field ($<10^{10}$ G) NS, similar as those in LMXBs. The observed
spectral features \citep{Shirey1999,Soleri2009} and kilo-Hertz
quasi-periodic oscillations \citep{Boutloukos2006} also lend strong
support to this conjecture. However, the orbit is shown to be
eccentric with $e=0.45$ \citep{Jonker2007}, suggesting that the
binary was recently subject to mass loss (and/or a kick) during the
formation of the compact star. Moreover, \citet{Heinz2013}
discovered a supernova remnant (with age less than 4600 yr)
surrounding Cir X$-$1, making it the youngest known  X-ray binary.
\citet{Jonker2007} associated the observed optical absorption-line
spectrum with the companion star of Cir X$-$1, suggesting that the
companion star could be a B5$-$A0 supergiant. However, the
possibility cannot be excluded that the spectrum originates in the
accretion disk, and in that scenario, the companion star is likely
to be a (sub)giant of mass around $0.4 M_\sun$. If it is the case
the conflicting properties (young age, eccentric orbit, and low
magnetic field) of Cir X$-$1 can be well accounted for by the NS-SS
phase transition scenario. The key point is that the supernova
remnant and the eccentric orbits were produced by the quark nova
when the accreting NS was converted to be a SS, rather than the
supernova that originally formed the NS.

No matter if our specific scenario is confirmed or ruled out by
future observations, the present data discussed in this paper will
provide useful constraints on the formation channels of these
peculiar systems, and there will be deep consequences for the
physics of strong interactions.

\begin{acknowledgements}
We are grateful to an anonymous referee for helpful and constructive
comments. This work was supported by the Natural Science Foundation
of China under grant Nos. 11133001 and 11333004, and the Strategic
Priority Research Program of CAS under grant No. XDB09010000.

\end{acknowledgements}

%\section{References}

\newpage

%\begin{figure}[h,t]
%\centerline{\includegraphics[angle=90,width=1.00\textwidth]{fig1.eps}}
%\caption{Flow chart of Monte Carlo simulation for orbital change. Input parameters are shown with their ranges. \label{figure1}}
%\end{figure}

\begin{figure}[h,t]
\centerline{\includegraphics[angle=0,width=1.00\textwidth]{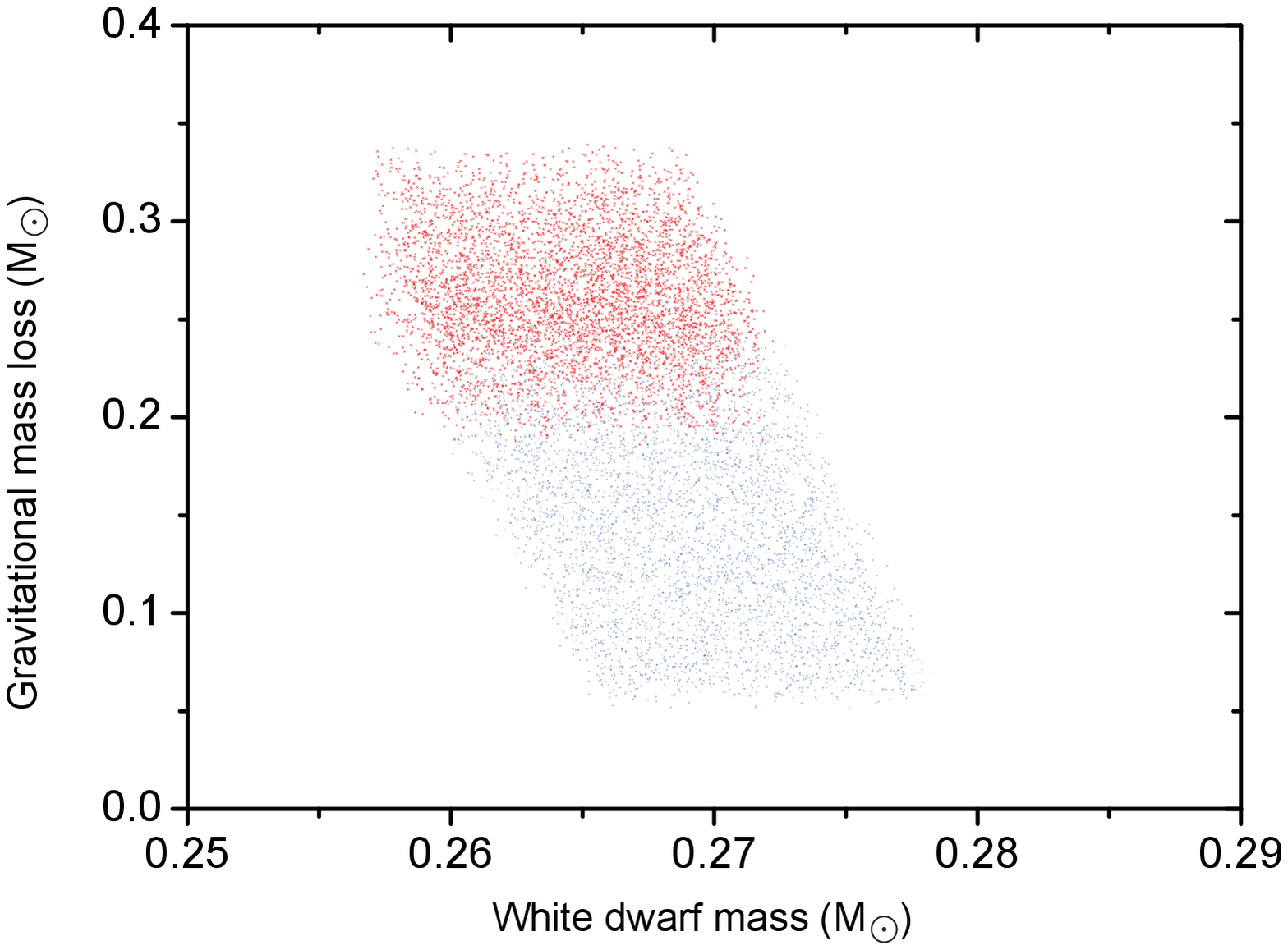}}
\caption{Required gravitational mass loss of the NS vs. the WD mass
in the phase transition scenario to reproduce the properties of PSR
J2234$+$06. The red and blue dots represent the cases for the
eccentricity in the range of $0.11-0.15$ and $0.027-0.11$,
respectively. \label{figure2}}
\end{figure}

\begin{figure}[h,t]
\centerline{\includegraphics[angle=0,width=1.00\textwidth]{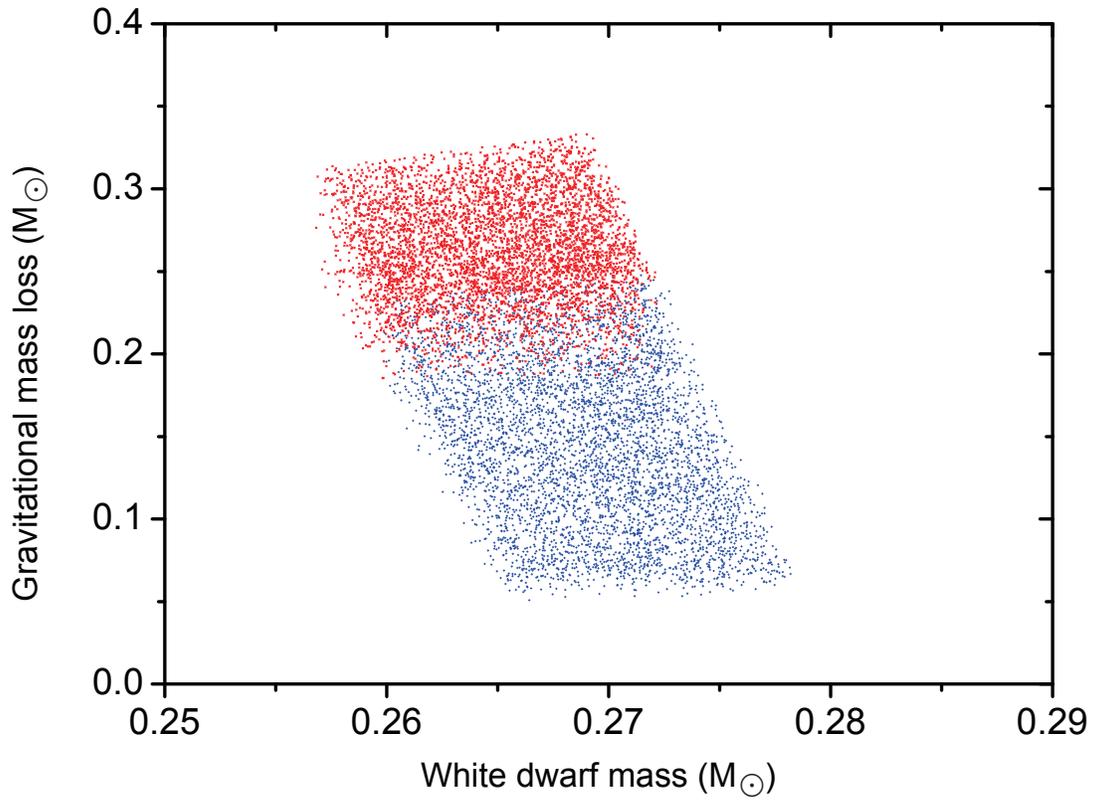}}
\caption{Same as Fig.~1 but with the mass function for PSR
J1946$+$3417. \label{figure3}}
\end{figure}

\end{document}